\documentclass[%
reprint,
superscriptaddress,
showpacs,preprintnumbers,
showkeys, 
 amsmath,amssymb,
 aps,
 pre,
]{revtex4-1}

\usepackage{graphics}
\usepackage{todonotes}
\usepackage{hyperref}

\begin{document}
\title{Collective excitability in a mesoscopic neuronal model of epileptic activity}
\author{Maciej Jedynak}
\email{maciej.jedynak@protonmail.com}
\affiliation{Departament de F\'isica, Universitat Polit\'ecnica de Catalunya (UPC), Colom 11, E-08222 Terrassa, Barcelona, Spain}
\affiliation{Department of Experimental and Health Sciences, Universitat Pompeu Fabra (UPF), Parc de Recerca Biom\`edica de Barcelona, Doctor Aiguader 88, E-08003 Barcelona, Spain}
\author{Antonio J. Pons}
\affiliation{Departament de F\'isica, Universitat Polit\'ecnica de Catalunya (UPC), Colom 11, E-08222 Terrassa, Barcelona, Spain}
\author{Jordi Garcia-Ojalvo}
\affiliation{Department of Experimental and Health Sciences, Universitat Pompeu Fabra (UPF), Parc de Recerca Biom\`edica de Barcelona, Doctor Aiguader 88, E-08003 Barcelona, Spain}

\begin{abstract}
The brain can be understood as a collection of interacting neuronal oscillators, but the extent to which its sustained activity is due to coupling among brain areas is still unclear.
Here we study the joint dynamics of two cortical columns described by Jansen-Rit neural mass models, and show that coupling between the columns gives rise to stochastic initiations of sustained collective activity, which can be interpreted as epileptic events.
For large enough coupling strengths, termination of these events results mainly from the emergence of synchronization between the columns, and thus is controlled by coupling instead of noise.
Stochastic triggering and noise-independent durations are characteristic of excitable dynamics, and thus we interpret our results in terms of collective excitability.
\end{abstract}

\maketitle

\section{Introduction}

Coupling provides a pervasive mechanism to organize populations of dynamical elements, via for instance synchronization and clustering \cite{Boccaletti:2002aa,Pikovsky:2003aa,Manrubia:2004aa}.
While in many cases coupling merely coordinates dynamical regimes that are already present in the isolated elements, in others it underlies the {\em emergence} of novel behaviors that would not exist in the absence of interaction between the elements \cite{De-Monte:2007aa,Zamora-Munt:2010aa}.
In the brain, examples of such emergent behavior exist at the microscopic scale of networks of neurons, in the form of, for instance, collective oscillations arising from a balance between excitation and inhibition \cite{Van-Vreeswijk:1996aa,Sancristobal:2014aa} and recurrent up/down dynamics \cite{Sancristobal:2016aa}.
Much less is known, however, about the emergent behavior of the brain at the mesoscopic level of coupled brain areas. 

Mesoscopic brain models often discretize spatially extended neuronal tissue and introduce a coupling strength that refers to the magnitude of interaction between nodes (typically neural mass models describing cortical columns \cite{Pons2010}) forming a discretised mesh or network.
In practical terms, this quantity is most often a multiplier of the output activity of an upstream node, before this output is fed to a downstream node. 
It is not clear, however, how to derive this coupling strength experimentally.
For this reason, in computational models the coupling strength is often used as a scan parameter \cite{Deco2013}.
Such variations of the coupling strength, with other parameters of an excitatory network kept constant, entail changes of the effective input received by the downstream nodes. 
In such conditions it becomes unclear to what extent the changes of dynamical properties of the system emerge from inter-node interactions and to what extent they simply follow from the intra-node dynamics driven by an increased effective net input. 

For example, \citet{Huang2011a} find, in a system of two interconnected neural mass modules, that an increase of coupling strength leads to vanishing of equilibria.
Nevertheless, they also show that a similar effect occurs in case of a single module, i.e. in the absence of inter-module coupling, where an increase of a constant input leads to Hopf and saddle-node-on-invariant circle (SNIC) bifurcations \cite{Grimbert2006} that result in the disappearance of the associated stable fixed points.
This SNIC bifurcation and the associated limit cycle are often used in theoretical studies of epilepsy \cite{Touboul2011}. 
For example, \citet{Goodfellow2016} study the generation of epileptic events in a system of coupled cortical columns described by the Jansen-Rit model, enriched with additional inhibitory processes and operating in an excitable regime. 
They observe that the net input to individual nodes, dependent on coupling and/or connectivity, is one of the factors contributing to occurrence of epileptic-like activity. 
Also in the context of epilepsy, \citet{Goodfellow2012} study the spreading of transient epileptic-like excitations on a lattice of interconnected Jansen-Rit modules, examining the role of coupling and observing that mixing of two oscillatory modes gives rise to transient excitations. 
The authors do not, however, quantify how coupling affects the durations of these excitations, nor do they consider noise, which is ubiquitous in the nervous system \cite{Faisal2008}.

Here we focus on the temporal properties of such transient excitations in a system of two coupled Jansen-Rit modules subject to white noise.
These transients arise due to a non-trivial feature of the Jansen-Rit model, namely the coexistence of two limit cycles, one of which displays quasiharmonic oscillations of frequency $\sim 10$~Hz, resembling the alpha brain activity, and the second of which displays spiky epileptic-like behavior (see for instance Fig. 2 in \cite{Jedynak2017}). 
The interplay between these two dynamics is physiologically relevant: in the epileptic brain the spreading of seizures may be gradual, therefore both types of oscillatory modes may coexist and interact. 

We study the dynamical properties of these excitation episodes by systematically varying the coupling strength between two reciprocally connected Jansen-Rit modules. 
In order to initiate excitations we subject the two modules to Gaussian white noise.
Importantly, we use compensated inputs when varying the coupling strength acting upon each module.
This prevents simple crossings through the bifurcation due to the increased net input caused by the increase of coupling between the modules.
In that way, regardless of the value of the coupling strength, the steady state (the stable node) of the system is always located equally far from the excitability threshold (the SNIC bifurcation), which allows us to focus on effects arising solely from the coupling between the modules.
As we show below, this coupling together with stochasticity gives rises to sustained collective activity that exhibits features characteristic of an excitable behavior, and therefore we term it {\em collective excitability}.
In particular, our results show a non-monotonic trend of the initiation rate of the excitation episodes as a function of the coupling strength. We also observe coupling-mediated synchronous terminations of these episodes, which bear resemblance to clinically observed epileptic activity \cite{Schindler2007}.  

\section{Coupled neural mass model}

\begin{figure}[htb]
   \centering
   \includegraphics[width=0.5 \textwidth]{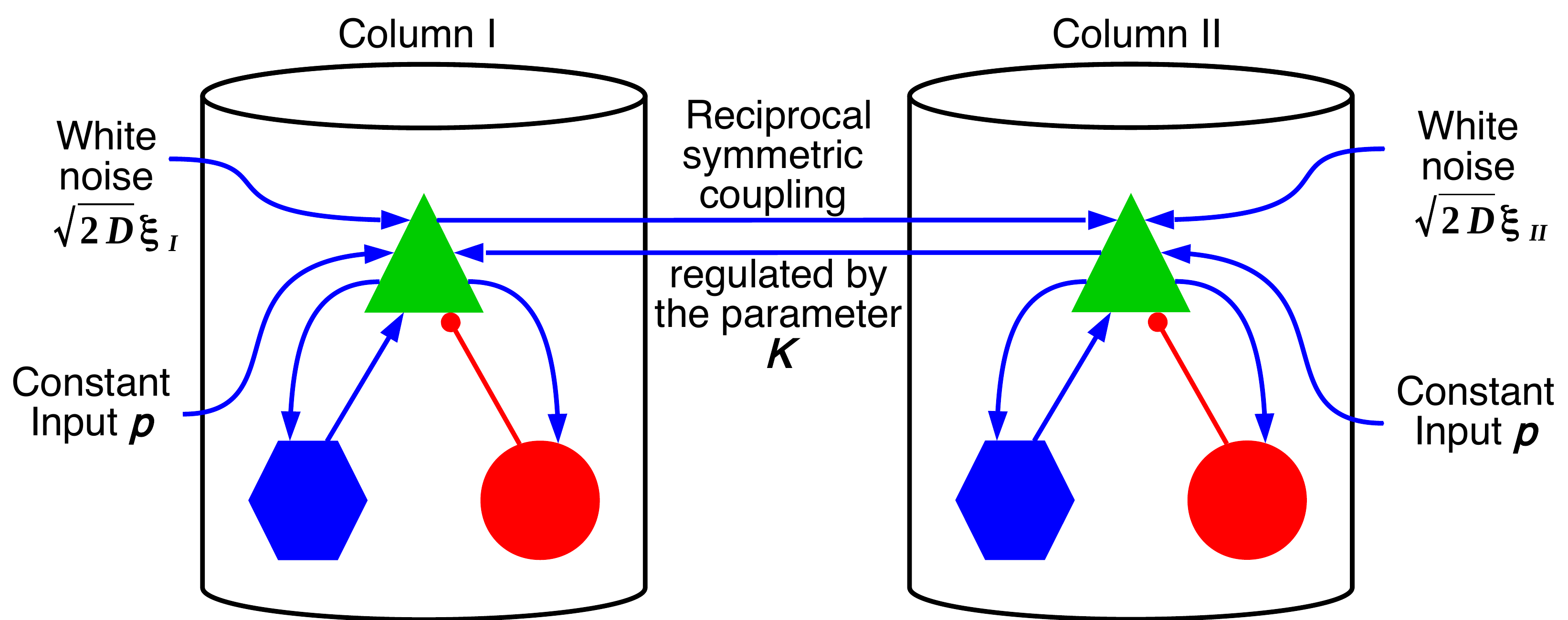}
	\caption{
	A system of two reciprocally coupled Jansen-Rit cortical columns. 
	Each column consists of three neuronal populations: pyramidal neurons (green triangles), excitatory interneurons (blue hexagons) and inhibitory interneurons (red circles). 
	Red (blue) connectors indicate inhibitory (excitatory) connections. 
	Each column is driven by three external signals acting upon the population of pyramidal neurons: 1) output from the interconnected column, 2) white noise, and 3) a constant signal.
	The two latter components stand for background activity of brain areas that are not modeled directly.
	} 
\label{fig:columns}
\end{figure}

We use a system of two coupled Jansen-Rit neural mass models \cite{Jansen1993, Jansen1995} driven by independent realizations of white noise. 
A scheme of inter- and intra-columns connectivity of this system is presented in Fig.~\ref{fig:columns}.
This figure shows that each of the cortical columns comprises three neuronal populations, consisting respectively of pyramidal neurons, excitatory interneurons, and inhibitory interneurons (see figure caption for details). 

The dynamics of these populations are governed by two transformations.
The first transformation is linear and converts presynaptic spiking activity $s_{\rm{in}}$ to postsynaptic membrane potential $y(t)$ (PSP), via the following convolution:
\begin{equation}  \label{eq:convolution}
    y(t) = \int_{0}^{\infty} h(t')s_{\rm{in}}(t-t')\mathrm{d}t',
\end{equation}
where $s_{\rm{in}}(t)$ is the time-dependent average firing rate of spike trains incoming to a population, $y(t)$ is the resulting PSP, and the $h(t)$ kernel describes the PSP response at the soma resulting from an impulse activation at the synapse. 
This kernel is zero for $t<0$, and otherwise it is given by the following expression for excitatory and inhibitory connections:
\begin{alignat}{2}  
	& h_{e}(t) && = Aate^{-at} \label{eq:exc_kernel}, \\
	& h_{i}(t) && = Bbte^{-bt} \label{eq:inh_kernel},
\end{alignat}
where $A$ and $B$ are the maximum excitatory and inhibitory PSPs, respectively, and $a$ and $b$ are reciprocals of the time constants of these responses. 
These constants follow from lumped contributions of all dilatory effects that include synaptic kinetics, dendritic signal propagation and leak currents. 
Eq.~(\ref{eq:convolution}) can be expressed, using Eq.~(\ref{eq:exc_kernel}), by the following differential form:
\begin{equation}  \label{eq:differential}
	\frac{d^2y(t)}{dt^2} +  2a\frac{dy(t)}{dt} +  a^2y(t) = Aa\cdot s_{\rm{in}}(t),
\end{equation}
Similarly, by using Eq.~(\ref{eq:inh_kernel}) one can find a corresponding differential representation of the inhibitory population dynamics.

The second transformation is nonlinear and converts the net membrane potential $y(t)$ to efferent firing rate $s_{\rm{out}}$.
It is given by the following sigmoid function:
\begin{equation} \label{eq:sigmoid}
	{\rm Sigm}(y) = \frac{2e_{0}}{1 + e^{r(\nu_{0} - y)}},
\end{equation}
where $y$ is the instantaneous net PSP (in general, time dependent). $s_{\rm{out}}(y)$, which is proportional to ${\rm Sigm}(y)$, is the resulting average firing rate of the spike train outgoing from the population. The response is defined by the maximum firing rate $2e_{0}$, the PSP $\nu_{0}$ for which a half maximal firing rate is reached, and the steepness (and thus nonlinearity) $r$ of this transformation. 

In case of the system presented in Fig.~\ref{fig:columns}, the transformations described above lead to a set of six second-order ordinary differential equations, which can be written in the following compact form:
\begin{eqnarray}
	\left(\frac{d}{dt} + a\right)^2 y^{i}_0(t) & = & Aa\;{\rm Sigm}[y^{i}_1 (t) - y^{i}_2(t)], \label{eq:JansenRit1} \\*
	\nonumber \left(\frac{d}{dt} + a\right)^2 y^{i}_1(t) & = & Aa\{p + K\;{\rm Sigm}[y^{j}_1 (t) - y^{j}_2(t)] + \\*
	&& + C_2 \;{\rm Sigm} [C_1 y^{i}_0(t)] + \sqrt{2D}\xi_i(t) \}, \label{eq:JansenRit2} \\*
	\left(\frac{d}{dt} + b\right)^2 y^{i}_2(t) & = & Bb\{C_{4}\;{\rm Sigm}[C_3 y^{i}_0(t)]\} \label{eq:JansenRit3}
\end{eqnarray}
where the tuple of indexes $(i,j)$ takes values \{(I,II),(II,I)\}, with 'I' and 'II' denoting the two Jansen-Rit modules (see Fig.~\ref{fig:columns}).
For each module, 
$y_0(t)$ is proportional to the excitatory PSPs induced on the two interneuron populations, $y_1(t)$ is the excitatory PSP induced on the population of pyramidal neurons, and $y_2(t)$ is the inhibitory PSP on that same population. 
As a consequence of these definitions, $y_1(t) - y_2(t)$ is the net membrane potential on the population of pyramidal neurons, which is assumed to be approximately proportional to the EEG measured close to the cortical column.
$\xi_{\rm I,II}(t)$ are random variables representing two independent Gaussian white noise realizations, each characterized by zero mean and correlation $\langle \xi_i(t) \xi_i(t')\rangle = \delta(t-t')$. 
The parameter $D$ represents the intensity of both noise terms. 

Following the existing literature \cite{Jansen1995} we use the following parameter values: $e_{0}=2.5~s^{-1}$, $v_{0}=6~{\rm mV}$, $r=0.56~{\rm mV^{-1}}$, $A=3.25~{\rm mV}$, $B=22~{\rm mV}$, $a=100~{\rm s^{-1}}$, $b=50~{\rm s^{-1}}$, $C_{1}=135$, $C_{2}=108$, $C_{3}=C_{4}=33.75$. For these parameters, the value of $p$ chosen here (see below) and in the absence of coupling, a single neural mass operates in either an excitable regime in which noise can produce isolated spikes, or in an oscillatory regime at a frequency of around 10~Hz, which can be identified as an alpha regime.

We integrated the system numerically using the stochastic Heun scheme \cite{toral}.
For each setting of $D$ and $K$ values we performed 10 simulations of duration 3600~s (one additional second was used to buffer the sliding window), furnished with different realizations of the stochastic processes, frozen for different parameters settings.
For each value of $K$ we chose $p$ in such a way that the system operates, in the absence of noise, at a distance $\Delta p=1~{\textrm{s}^{-1}}$ below the excitability threshold, given by the SNIC bifurcation. Details on the localization of the bifurcation point are given in the Appendices.

\section{Transient collective excitations}

\begin{figure}[htb]
   \centerline{\includegraphics[width=0.5 \textwidth]{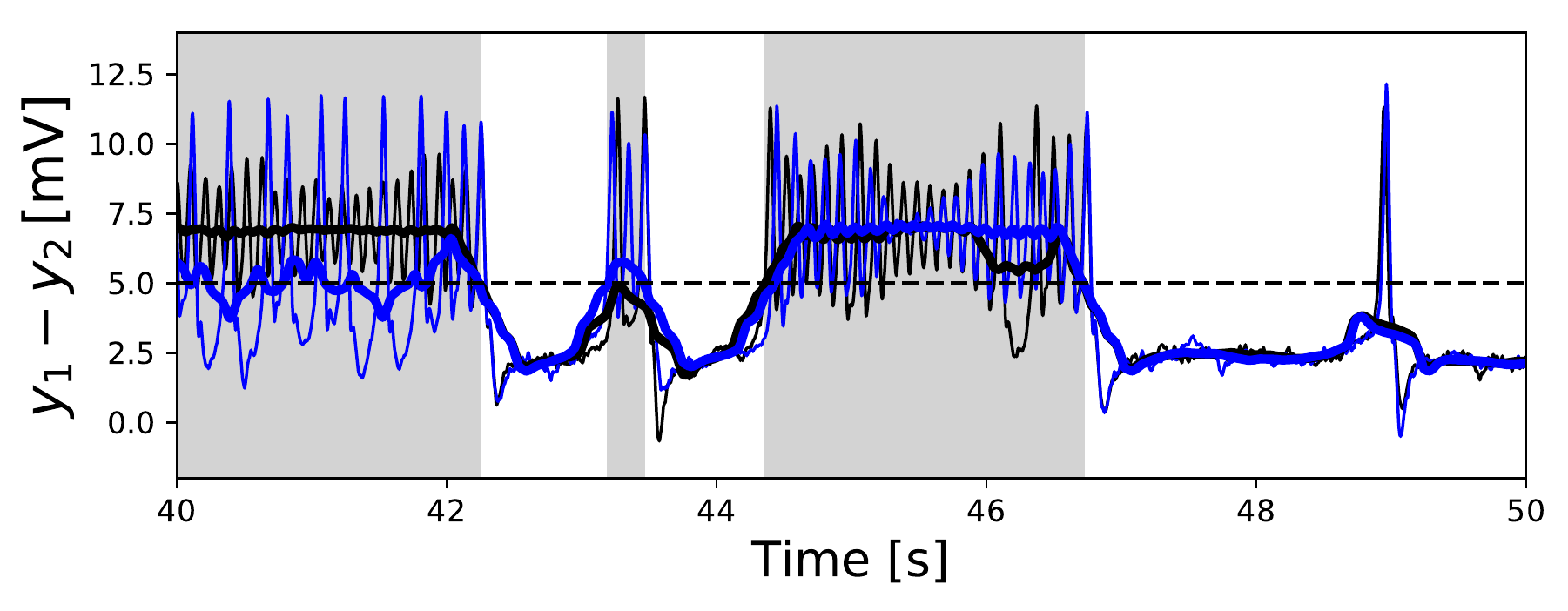}}
	\caption{
	Typical dynamics of the two columns, in terms of the net PSP signal $y_1(t) - y_2(t)$, marked by thin black and blue lines.
	Running averages of these signals obtained with a sliding window of length 0.5~s are shown by thick black and blue lines.
	The threshold $T$ is shown as a dashed line, and periods classified as prolonged activity are represented by the grey background.
	Here coupling strength is $K=10$, the intensity of white noise is $D=0.5~{\textrm{s}^{-1}}$, and $p$ is adjusted to $K$ according to the methodology introduced in text. The rest of the parameters are given in the text.   
	} 
\label{fig:mmtt}
\end{figure}

The typical dynamics of the system of two coupled neural mass oscillators described above is shown in Fig.~\ref{fig:mmtt} for moderate coupling strength, where the thin black and blue lines represent the output signals of the two columns ($y^{i}_1 - y^{i}_2, i \in \{{\rm I,II}\}$).
The time series shows periods of sustained activity (marked by a grey background), interspersed by regions in which the two columns are quiescent or exhibit at most single spikes (white background). 
The transient episodes do not occur due to slow changes of a system parameter that would lead to hovering over the bifurcation \cite{LopesDaSilva2003, Baier2012}, but are instead caused, as we will see below, by stochastically-initiated complex interactions between the two oscillatory modes.

The regions of sustained activity and quiescence can be identified by calculating running averages of the two signals within a sliding window of length $W$ (thick black and blue lines).
Spans of time in which at least one of these averages is above the threshold $T$ are considered excitation periods. 
For the values of $W$ and $T$ chosen ($W=0.5$~s and $T=5$~mV), the result of the classification agrees with eye inspection. 
Pure epileptic-like single spikes are not considered excitation episodes, because we are interested in activity that can be transiently self-sustained and can therefore lead to longer excitation periods.
To emphasize this exclusion of single spikes, we use the term ``prolonged excitation transients'' to refer to this regime.
Note that the synchronous spiking occurring around $t=49$~s is not considered prolonged activity, whereas the short episode appearing between $t=43$~s and $t=44$~s is considered as such.

\section{Influence of coupling in transient dynamics}

\begin{figure}[htb]
   \includegraphics[width=0.5 \textwidth]{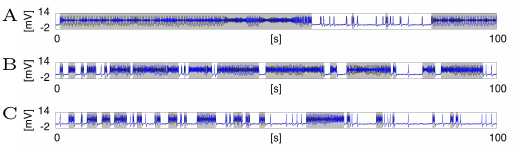}
	\caption{
	Typical time courses obtained for three different values of the coupling strength.
	100 seconds of activity of the model are shown for $K=5$ (A), $K=10$ (B) and $K=15$ (C).
	Periods classified as prolonged activity are represented by a gray background.
	In all cases the noise intensity $D$ was set to $0.5~{\textrm{s}^{-1}}$, and the system was operating $\Delta p=1~{\textrm{s}^{-1}}$ below the excitability threshold set by the SNIC bifurcation.
	} 
\label{fig:tt}
\end{figure}

We next examine the role of coupling in the prolonged excitation transients described above.
To that end, we vary the coupling strength $K$ while modifying the input $p$ such that the system of two coupled modules operates at a fixed distance ($\Delta p=1~{\textrm{s}^{-1}}$) from the excitability threshold.
Figure~\ref{fig:tt} shows examples of time courses obtained for three different values of coupling $K$, and in the presence of noise of intensity $D=0.5~{\textrm{s}^{-1}}$. 
The data shows that the durations of both the prolonged excitation transients and the quiescent periods depend strongly on the coupling strength between the columns. 
In particular, for small coupling (panel A, $K=5$) the system does not change the state as frequently as for larger coupling strengths (panels B and C, $K=10$ and $15$, respectively). 
In contrast, for intermediate values of $K$ (panel B, $K=10$) the system switches to the excited state most readily.
Next, we studied these effects by systematically varying $K$ and performing long simulations that allowed us to compute the initiation and termination rates of the prolonged excitation transients with high statistical accuracy. 

The initiation rate is defined as the total number of prolonged excitation transients divided by the total duration of the quiescent state (noise-driven fluctuations around the steady state).
In turn, the termination rate is defined as the total number of terminations of the sustained events divided by their total duration. 
These quantities are not computed for each realization of the noise separately, but rather within each set of parameters once for all realizations considered together. 
This procedure minimizes effects related to the finite time of the simulation, and at the same time involves various realizations of stochastic processes.

\begin{figure}[htb]
   \includegraphics[width=0.5 \textwidth]{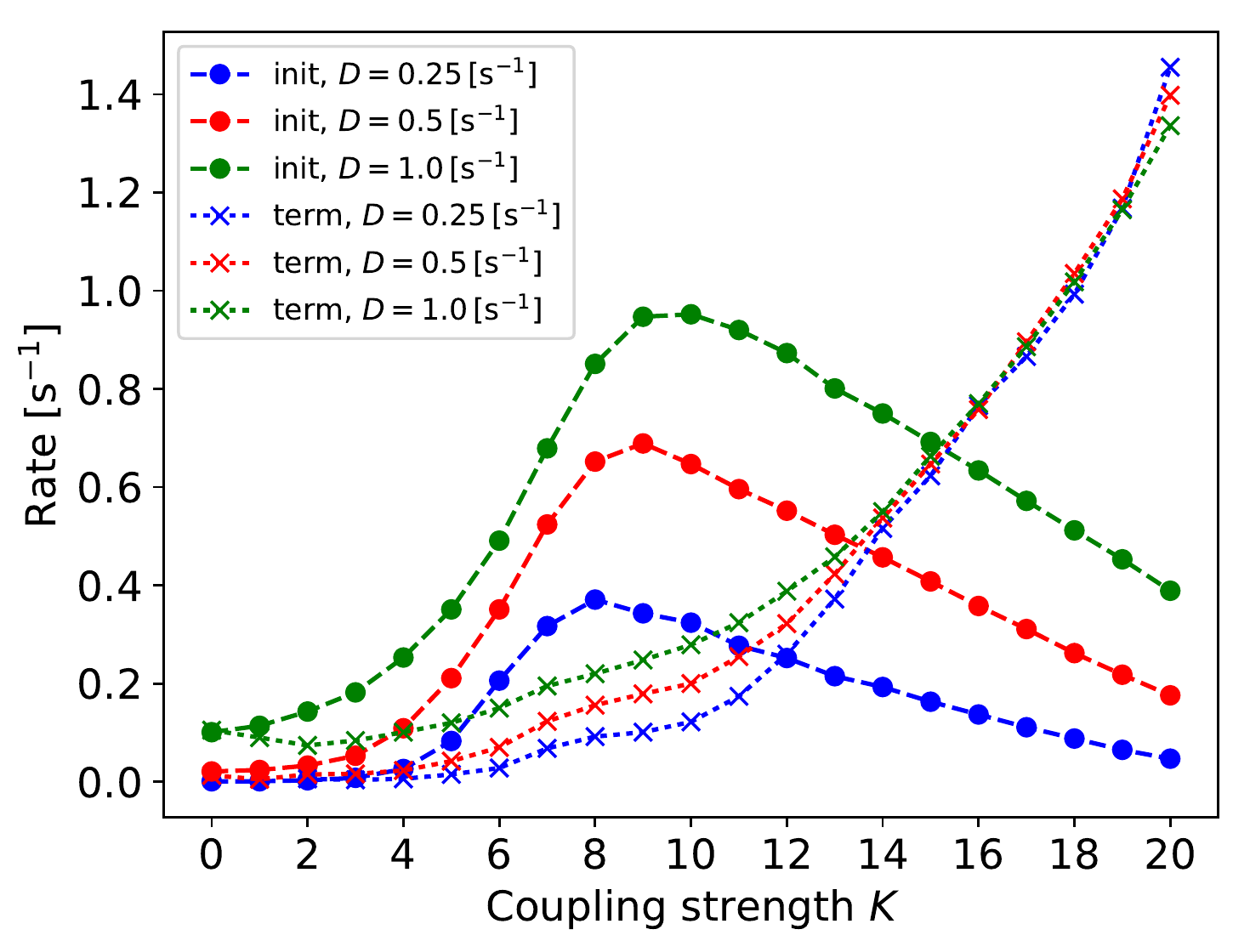}
	\caption{
	Dependence of the initiation and termination rates on coupling strength and noise intensity.
	Initiation (`init', circles) and termination (`term', crosses) rates for discrete values of the coupling strength $K$ are shown for three different intensities of the driving noise: $D=0.25~{\textrm{s}^{-1}}$ (blue), $D=0.5~{\textrm{s}^{-1}}$ (red), $D=1~{\textrm{s}^{-1}}$ (green). Dashed and dotted lines are plotted to guide the eye. 
	For the lowest noise intensity ($D=0.25~{\textrm{s}^{-1}}$) and two lowest coupling values ($K=1,2$) the initiation rate was too low to gather sufficient number of episodes to measure termination rate. 
	Therefore the two lowest points of termination rate for lowest noise intensity are considered outliers and are not plotted.
	} 
\label{fig:rates}
\end{figure}

Figure~\ref{fig:rates} shows how the initiation and termination rates of the prolonged excitation transients depend on the coupling strength $K$. 
The results are presented for three different intensities of the noise: $D=0.25~{\textrm{s}^{-1}}$, $0.5~{\textrm{s}^{-1}}$ and $1~{\textrm{s}^{-1}}$ (blue, red and green, respectively).
This figure demonstrates that for low values of $K$, both the initiation and termination rates are relatively low, indicating rare transitions between quiescent and excited states. 
$K=0$ corresponds to the case of two separated modules, and shows that in that case transitions between the states are rare, occurring only for sufficiently high noise intensities (green). 
For low noise intensity (blue) and low coupling $K$ the effect is virtually absent, while it emerges for increasing $K$.
The fact that coupling between the modules is required for excitations to occur indicates that this behavior is a collective phenomenon. 

The increase of $K$ leads to a moderate increase of termination rates and a rise of initiation rates, which reach a maximum for intermediate $K$ ($K=8$ for $D=0.25~{\textrm{s}^{-1}}$, $K=9$ for $D=0.5~{\textrm{s}^{-1}}$ and $K=10$ for $D=1~{\textrm{s}^{-1}}$). 
For larger $K$ values the initiation rates decay, whereas termination rates increase more rapidly. 
For large coupling the termination rate is basically independent of the noise intensity, showing that in that regime it is coupling, not noise, that plays a major role in the termination of activity. 
In other words, when $K$ is high, the prolonged excitation transient regime is hardly susceptible to noise. 
The concurrence of a noise-dependent initiation and noise-independent termination is a typical behavior of an excitable system \cite{Lindner2004}. 
In this system, large enough coupling induces excitable behavior at the level of two columns, which thus constitutes a collective excitable effect. 
For smaller couplings ($K<15$ in Fig.~\ref{fig:rates}), the termination rate depends on noise intensity, which is illustrated by the separation of dotted lines in Fig.~\ref{fig:rates}, and resembles other types of excitatory dynamics, such as the one observed in the calcium-mediated activity of cardiac myocytes \cite{Song2017}.

The trend of the initiation rate presented in Fig.~\ref{fig:rates} may also be explained as follows. 
Low $K$ values hamper recruitment between the columns: when one column spikes due to the stochastic perturbation, it is less likely to excite the other one, as long as $K$ remains low. 
For larger $K$ values, in contrast, a spike in one column entails a significant perturbation applied to the other column, and allows it to also leave the quiescent state. 
In the limit of very large coupling the two columns may display the same behavior -- they synchronously spike and simultaneously (and quickly) fall to the refractory period followed by the quiescent state (see again the isolated spike around $t=49$~s in Fig.~\ref{fig:mmtt}).
This does not correspond to an initiation since the activity is not sustained.
Intermediate $K$ values, in contrast, allow for the mixing of oscillatory modes; although the recruited column leaves the quiescent state, the columns do not fully synchronize and as a result at least one of them sustains activity by moving to the attractor of alpha oscillations, which lacks refractory period. 
These oscillations yield an increased output that is fed to the other column, which can again go active after leaving the refractory period, and a prolonged excitatory activity may develop. 
This effect explains why the initiation rate peaks for an intermediate coupling in Fig.~\ref{fig:rates}. 
The location of the peak in Fig.~\ref{fig:rates} varies for different noise intensities $D$, which can be explained by the fact that larger noise requires higher coupling strength for complete synchronization. 

\section{Synchronous termination of the transient excitations}

\begin{figure}[htb]
   \includegraphics[width=0.5 \textwidth]{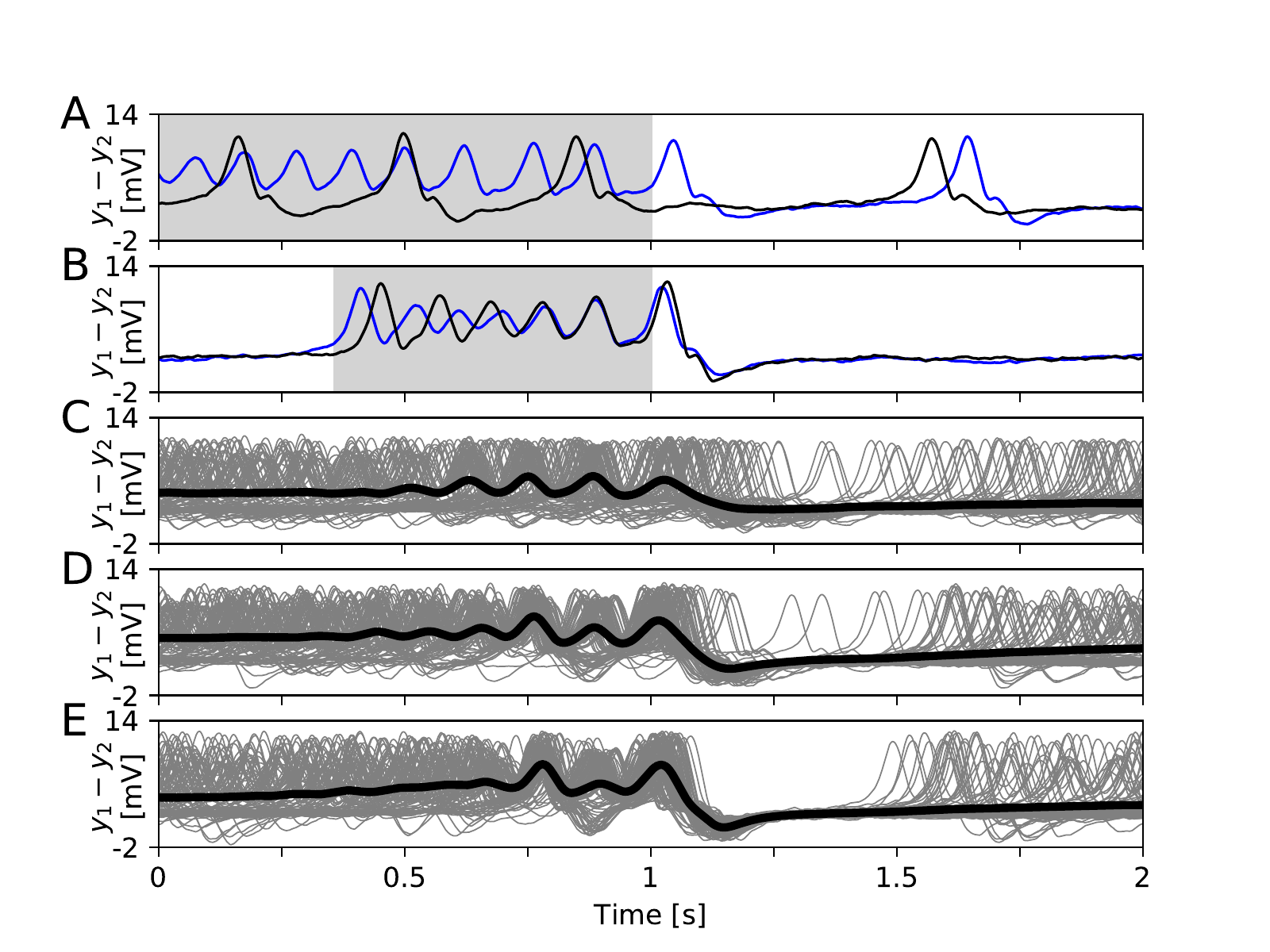}
	\caption{
	Terminations patterns of prolonged excitation transients. 
	Panels A and B show typical time courses of two cortical columns (black and blue lines) for $K=5$ and $K=15$, respectively.
	Periods of prolonged excitation transients are marked in grey.
	Panels C, D, E show averaged time courses from both columns (think black lines) and from all prolonged excitation transients for $K=5, 10, 15$, and obtained from averaging $2.5\cdot10^{3}$ , $11\cdot10^{3}$ and $18\cdot10^{3}$ individual time courses, respectively.
	The thin grey lines in the three panels represent are 100 typical time courses. 
	In all five panels, the point of excitation termination (identified by our classification algorithm) has been shifted to the middle of the plot (corresponds to $t=1$~s). 
	The results were obtained for $D=0.5~{\textrm{s}^{-1}}$.
	} 
\label{fig:avg}
\end{figure}

Why is the termination rate independent of noise for large enough coupling strength?
Figure~\ref{fig:avg} shows the activity of the system around the termination of prolonged excitation transients for different coupling values. 
The figure shows that for low coupling ($K=5$, panel A), the events terminate when one of the two columns falls stochastically back to the quiescent state, and thus termination is unsynchronised between the two columns.
Therefore, in that regime termination is mainly dominated by noise.
In contrast, for higher $K$ it occurs rather due to synchronization effects: Fig.~\ref{fig:avg}B shows that for high coupling ($K=15$) the synchrony between the two columns gradually develops, until a simultaneous drop of activity terminates the excitation period.
Although this gradual development of synchrony does not have to be present in every excitation transient, the synchronous termination is prevalent for higher coupling. 

The influence of $K$ on the synchronous termination is shown in Figs.~\ref{fig:avg}C-E, where the black line represents the average over both columns and all excitation transients for increasing coupling levels.
The grey lines shown in these panels are 100 typical time courses of individual columns.
They converge most strongly to the averaged time course (black line) after the termination ($t=1$~s) in the case of the largest  coupling level (Fig.~\ref{fig:avg}E).
Also in that case, the kink corresponding to the refractory period of the terminating spike is most strongly pronounced (dip in the black line after $t=1$~s). 
Stronger coupling promotes this synchronous termination, consistent with the faster buildup of synchronization patterns for $K=15$ (black line in Fig.~\ref{fig:avg}E) than for $K=5$ (black line in Fig.~\ref{fig:avg}C), as well as with the growth of the termination rate with $K$ presented in Fig.~\ref{fig:rates}.


\section{Discussion}

Here we have introduced the concept of collective excitability in a mesoscopic brain model, where prolonged excitation transients arise due to interactions between two connected modules, and terminate via synchronization buildup.
Our results could be applied to neurodynamical pathologies such as epilepsy, which is primarily characterized by excessive synchrony, with synchronization increasing during the development of seizures \cite{Truccolo2011a,Khambhati2014}.
Consistently with our findings, it has been proposed that epileptic seizures imply synchronization, and synchronization leads to seizure termination \cite{Majumdar2014}.
Termination of seizures has been shown to exhibit signatures of a {\em catastrophic transition} \cite{Kramer2012a} and occurs due to simultaneous entry into the refractory period, which has been hypothesized to be the scenario underlying a synchronous termination of seizures \cite{Schindler2007}. 
Both the ``catastrophic transition'' and the refractoriness are featured by our model.
Our results are also consistent with the observation that the locally simultaneous termination of seizures correlates with the coherent spike-and-wave activity \cite{Proix2017}. Moreover, our findings suggest that a common driving delivered to interconnected cortical columns (e.g. by brain stimulation) may lead to the full synchronization and thus seizure termination. This links our work to dynamical studies of stimulus driven epileptic seizure abatement \cite{Taylor2014}.

Given the large levels of synchronization that characterize the phenomenon, theoretical studies of epilepsy often involve a collapse of the spatial extension of the neuronal tissue, reducing the model of the brain from a network to a single node (a single cortical column in our case) \cite{Breakspear2006, Roberts2012}.
Our study deals with a dynamical regime in which this collapse cannot be performed, since the prolonged transient excitation episodes discussed here require the interaction between (at least) two distinct columns to sustain the dynamics, through the mutual activation of the two columns operating in distinct dynamical modes.
It is also worth mentioning that it is necessary to break the symmetry of the compound system in order for transient excitations to arise. Here, symmetry breaking is provided by the independence of two realizations of the stochastic processes affecting the two modules that form the system.

Our mixed oscillatory modes form complex transients, which occasionally may consist of solely healthy activity (alpha oscillations, see Fig.~\ref{fig:mmtt} around 45~s). These periods, however, are preceded and followed by activity containing also epileptic-like spiking. Such intermittent behavior is consistent with experimental observations concerning epilepsy \cite{Goodfellow2013}.
The recruitment to alpha activity in a column by an epileptic-like spike incoming from the other column might be facilitated by the fact that the columns are operating in a bistable regime, where alpha oscillations are one of the possible attractors of the system.

Our findings also have implications for generic studies of synchronization and collective self-organization. 
First, we have demonstrated a system which does not require slow changes to its parameters values in order to exhibit activity of durations greatly exceeding temporal timescales of these parameters.
Second, each one of our coupled elements may be considered a low-pass filter acting upon uncorrelated white noise, effectively rendering it temporally correlated before it is fed to the other column. This is consistent with previous work \cite{Jedynak2017} showing that temporal correlations of the driving stochastic input facilitate excitations in this model. It would be interesting to study if other coupled stochastic systems exhibit this connection between collective dynamics and temporal processing.

\appendix

\section{Bifurcation analysis}
\begin{figure}[htb]
   \centering
   \includegraphics[width=0.5 \textwidth]{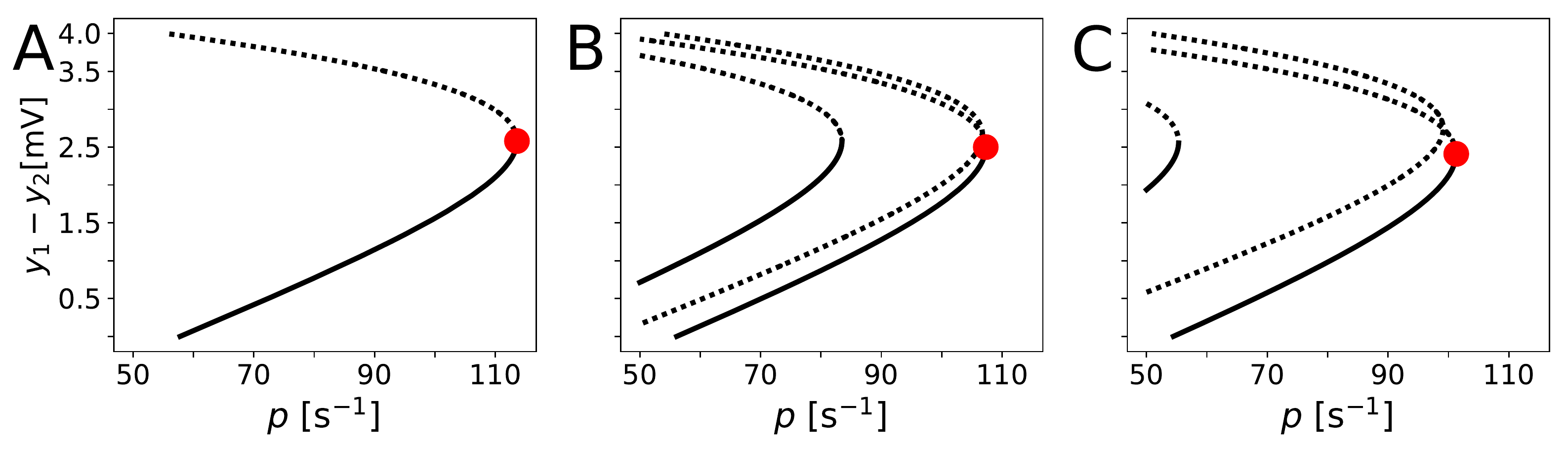}
	\caption{
	Bifurcation structure of the model around the SNIC bifurcation for the system of two coupled Jansen-Rit modules. 
	The three panels A, B, and C demonstrate invariant sets for the coupling strength $K$ equal 0, 10, and 20, respectively. 
	Continuous (dashed) lines mean stable (unstable) equilibria. 
	The red dot marks location of the SNIC bifurcation. 
	} 
\label{fig:snics}
\end{figure}

In order to localize the excitability threshold of the system governed by Eqs.~(\ref{eq:JansenRit1})-(\ref{eq:JansenRit3}), we performed a bifurcation analysis \cite{Ermentrout2002} of the deterministic part of this system.
We focused our attention on the location of the SNIC bifurcation, which bounds from below, in the absence of coupling, a regime of bistability between the two limit cycles \cite{Grimbert2006, Spiegler2010}. 
Figure~\ref{fig:snics} shows this SNIC bifurcation for three values of the coupling strength $K$ along with the associated stable (continuous line) and unstable (dashed line) branches. 

As mentioned in the main text, in the numerical simulations we choose $p$ in such a way that for each value of $K$ the system operates a fixed distance $\Delta p=1~{\textrm{s}^{-1}}$ below the excitability threshold (the SNIC bifurcation).
For example, for $K$=0, the SNIC bifurcation occurs at $p_{\textrm{snic}}=113.58~{\textrm{s}^{-1}}$, therefore we choose the constant part $p$ of the input to be equal to $p=112.58~{\textrm{s}^{-1}}$, whereas for $K=10$, $p_{\textrm{snic}}=107.3~{\textrm{s}^{-1}}$, thus we set $p$ to $106.3~{\textrm{s}^{-1}}$.
Since the location of this SNIC bifurcation marks the onset of epileptic activity, it is of high importance in theoretical studies on epilepsy \cite{Touboul2011}.
Figures~\ref{fig:snics}A-C demonstrate that for growing $K$, $p_{\textrm{snic}}$ decreases. 
Next we introduce a method of tracking this location in a system of coupled modules.

\section{Saddle-node bifurcation tracking}

We now introduce a simple method of tracking a loss of stability due to a saddle-node (SN) bifurcation in a compound system.
We also validate this method with a codimension-2 bifurcation analysis on the $K-p_{\textrm{snic}}$ plane performed with XPPAUT \cite{Ermentrout2002}. 
Such a loss of stability might be of special importance, e.g. when the SNIC bifurcation leads to a ``catastrophic'' onset of an epileptic-like limit cycle. 
This is a case of the Jansen-Rit model discussed in this paper. 
The ``perturbation convergence'' method presented in what follows was developed and tested on that model.

\begin{figure}
   \centering
   \includegraphics[width=0.5 \textwidth]{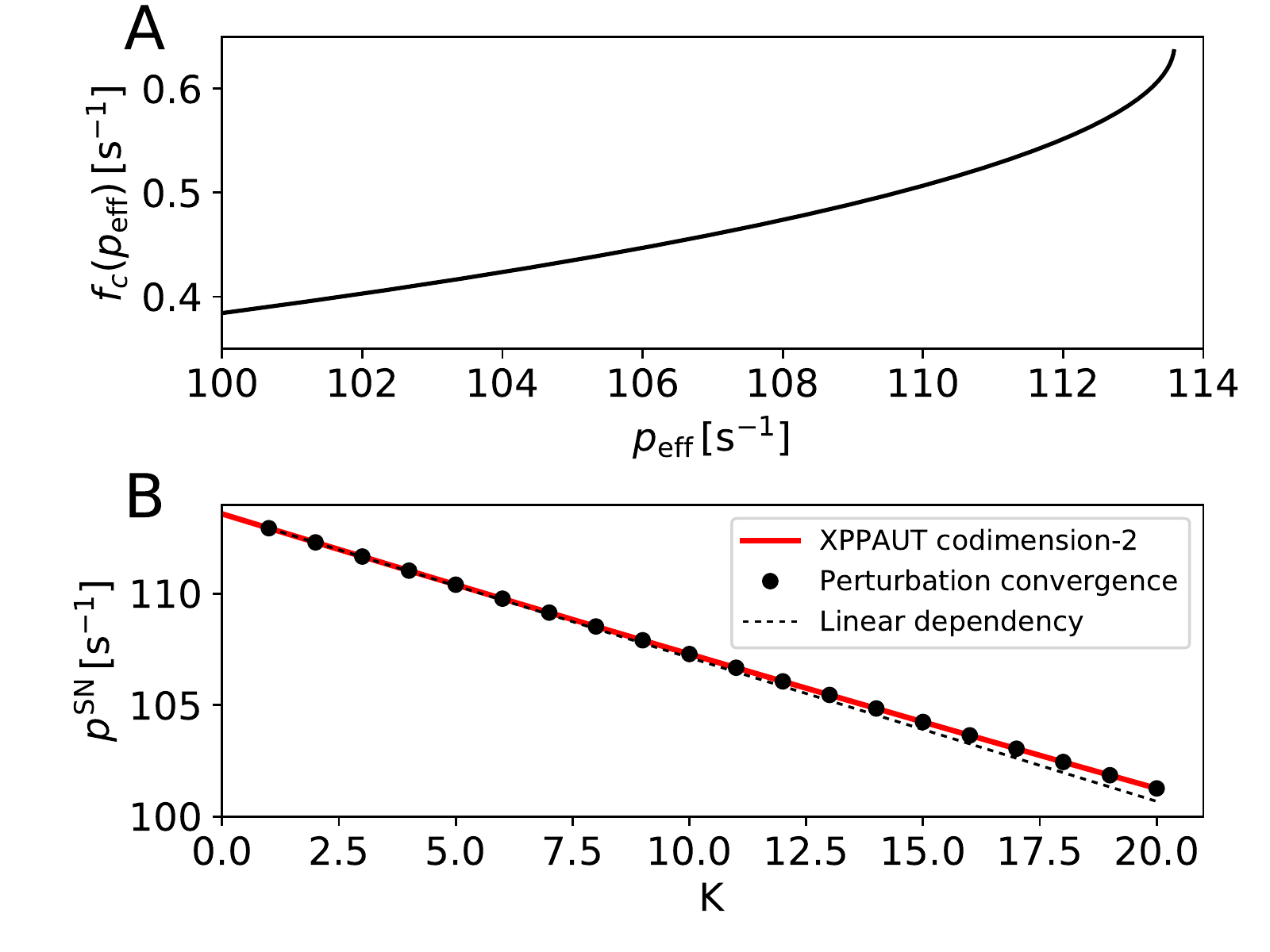}
	\caption{Panel A: coupling function characterizing the interaction between two columns. Panel B: location of the saddle-node bifurcation in the system of two connected Jansen-Rit modules computed with XPPAUT (red line) and with the perturbation convergence method. The approximated linear dependency is marked by the dotted line.} 
\label{fig:supp}
\end{figure}

The solid line in Fig.~\ref{fig:snics}A shows a stable solution (a node) for the uncoupled system of two Jansen-Rit columns exposed to an external constant driving $p$. 
This system is described by the deterministic part of Eqs.~(\ref{eq:JansenRit1})-(\ref{eq:JansenRit3}) with $K=0$, and has the properties of a single column. 
When coupling between the two columns increases, however, the SNIC bifurcation moves to the left (compare the locations of the red dot in panels A-C in Fig.~\ref{fig:snics}).
If $p$ is kept constant within the stable branch (continuous line in Fig.~\ref{fig:snics}A), the distance between the working point of the system and the SNIC decreases as the coupling strength increases, and the output of the system (read from the Y axis of Fig.~\ref{fig:snics}) increases.
This output, however, entails a further increase of the input fed to the interconnected column, which again provokes an increase of its output, and a similar influence is exerted by the second column upon the first. 
This transient process continues updating the effective input received by the columns $p_{\rm eff}$. 
Existence of the steady state (understood as the stable node below the SNIC bifurcation) requires that $p_{\rm eff}$ is finite, thus updates to $p_{\rm eff}$ must decrease in time. 
In order to express this condition formally, let us define a coupling function $f_{\rm c}(p_{\rm eff})$, which multiplied by $K$ converts the input $p_{\rm eff}$ delivered to column $i$ into this column's output.
\begin{equation} \label{eq:fc}
	f_{\rm c}(p_{\rm eff}) = {\rm Sigm}[y^*_1(p_{\rm eff}) - y^*_2(p_{\rm eff})],
\end{equation}
where $y^*_{1,2}(p_{\rm eff})$ denotes the value of the $y_{1,2}$ state variables in the steady state corresponding to $p_{\rm eff}$ and to the stable branch associated with the SN bifurcation.
Due to the symmetry of the system, the introduced dependencies are valid for both columns.
We therefore dropped for simplicity the index $i$ from the superscript of the $y$ state variables.
Furthermore, since the described method does not require that the tracked bifurcation is a global SNIC, but in general it can be a saddle-node as well, we from now on will refer to this bifurcation with the `SN' abbreviation.
For analytical implicit expressions for $y^*_{1(2)}(p_{\rm eff})$ see e.g. \cite{Huang2011a, Garnier2015}.
The $f_{\rm c}(p_{\rm eff})$ function is shown in Fig.~\ref{fig:supp}A.
By means of this function, we can express $p_{\rm eff}$ in the steady state as a sum of the external constant driving $p$ and the coupling term:
\begin{equation} \label{eq:steady}
	 p_{\rm eff} = p + Kf_{\rm c}(p_{\rm eff})
\end{equation}
A necessary condition for stability of the steady state of two coupled modules can be then written as:
\begin{equation}
	K \left.\frac{df_{\textrm{c}}(x)}{dx}\right|_{x=p_{\rm eff}} \leq 1
\end{equation}
If this condition is not fulfilled, a perturbation $\Delta p > 0$ of the external driving entails an increase of the coupling term exceeding $\Delta p$, which for monotonically increasing $f_{\textrm{c}}$ leads to divergence of $p_{\rm eff}$.
Thus, for monotonically increasing and differentiable $f_{\textrm{c}}$ and for $K > 0$, we expect the disappearance of the stable solution (SN bifurcation) at $p^{\rm SN}_{\rm eff}$ such that: 
\begin{equation} \label{eq:loss}
	\left.\frac{d f_{\textrm{c}}(x)}{dx}\right|_{x=p^{\rm SN}_{\rm eff}} = \frac{1}{K}
\end{equation}
from where Eq.~(\ref{eq:steady}) allows to find the corresponding external driving $p^{\textrm{SN}}$ (X coordinate of the bifurcation points, marked with red dots in Fig.~\ref{fig:snics}). 
Therefore, it is enough to know the coupling function $f_{\textrm{c}}$ and its derivative in order to quickly find the location of the SN bifurcation for an arbitrary $K$. 
Figure~\ref{fig:supp}B shows the location of the SN bifurcation of two coupled Jansen-Rit modules for increasing coupling, computed with the perturbation convergence method (black dots), along with the result of the codimension-2 analysis performed with XPPAUT (red line). 
In order to emphasize a slight deviation from the linear dependency, we plot it in a form approximating Eq.~(\ref{eq:steady}): $p^{\textrm{SN}} =  - Kf_{\rm c}(p^{\rm SN}_{\rm single}) + p^{\rm SN}_{\rm single}$ (dotted line), where $p^{\rm SN}_{\rm single} \sim p^{\rm SN}_{\rm eff}$ is where the SN bifurcation occurs for a single Jansen-Rit module.
Finally, let us note that by substituting $K$ in Eqs.~(\ref{eq:steady}),~(\ref{eq:loss}) by $K (N-1)$ we find the location of the loss of stability for a fully bidirectionally coupled network of $N$  Jansen-Rit modules.
Note that this method would not apply to instabilities arising from foci, because in that case, when relaxing to the focus, the system transiently exceeds the steady state value and therefore we can not apply the simple reasoning of the one-sided $p_{\rm eff}$ convergence.

\bibliography{bibliography}
\end{document}